\documentclass{moriond}

\def\gev{GeV/${c^2}$}
\def\tev{TeV/${c^2}$}
\def\csunit{cm${^2}$}

\def\be{\begin{equation}}
\def\ee{\end{equation}}
\def\bea{\begin{eqnarray}}
\def\eea{\end{eqnarray}}

\newcommand{\Photo}{}

\begin{document}
\vspace*{4cm}
\title{Constraints on Light WIMPs from SuperCDMS}

\author{A.~J.~Anderson\\on behalf of the SuperCDMS Collaboration}

\address{Massachusetts Institute of Technology, Department of Physics, Cambridge, MA 02139, USA}

\maketitle\abstracts{The SuperCDMS experiment searches for weakly interacting massive particles (WIMPs) using cryogenic germanium detectors that measure ionization and phonon energy. Several direct searches for WIMPs have recently reported excesses of events above their background expectations, which could be interpreted as WIMPs with masses in the 8-20~\gev~range. The excellent intrinsic signal-to-noise of SuperCDMS detectors and their powerful background rejection make SuperCDMS an ideal experiment to further test these signal hints. We present the results of two recent SuperCDMS analyses targeting this mass range: the first using Luke-Neganov amplification of the ionization signal to lower the effective energy threshold, and the second using multivariate methods to optimize background rejection in a larger 577~kg-d exposure.}

%We discuss a special operational mode of SuperCDMS, called CDMSlite, which lowers the effective energy thereshold to 170~eVee to probe WIMPs light as 4~\gev. We also present an analysis of 577~kg-d of exposure using boosted decision trees, which sets an upper limit on the spin-independent WIMP-nucleon cross section of $1.2 \times 10^{-42}$~\csunit~at 8~\gev.}

There is overwhelming observational evidence from astrophysics and cosmology that 85\% of the universe is in the form of non-luminous, non-baryonic, dark matter whose particle nature is unknown. Weakly interacting massive particles (WIMPs) are a well-motivated class of dark matter candidates, which persist as thermal relics from the early universe and can be detected by searching for keV-scale nuclear recoils in terrestrial experiments. Experiments worldwide are currently searching for WIMPs using a variety of techniques that probe masses between 1~\gev~and 10~\tev.

\section{The Light WIMP Scenario}
Well-motivated supersymmetric extensions of the standard model provide WIMP candidates favoring masses $>50$~\gev~when global fits are performed using data from colliders and cosmology.\cite{Buchmueller2013,Cotta2013,Strege2014} The absence of experimental signals of supersymmetry, however, has motivated the development of alternative models which accommodate WIMPs over a much larger mass range. Models such as asymmetric dark matter,\cite{Kaplan1992,Kaplan2009} in which the dark matter has an asymmetry of matter over antimatter that sets its relic density, and some dark photon models,\cite{Hooper2012} can accommodate WIMPs with significantly lighter masses in the range 1-10~\gev, a regime that can still be probed by existing direct detection experiments.

Complementing the theoretical possibility of light WIMPs, the DAMA/LIBRA,\cite{DAMA2013} CRESST,\cite{CRESST2012} CoGeNT,\cite{CoGeNT2013} and CDMS II (Si)\cite{CDMSSi2013} experiments have recently reported excesses of events over their background expectations. If interpreted as a signal, these excesses would correspond to WIMPs with masses in the range 8-20~\gev. There is considerable tension, however, with null results from CDMS II (Ge),\cite{CDMSGe2011} XENON10,\cite{XENON102011,XENON102011err} XENON100,\cite{XENON1002012} and more recently LUX.\cite{LUX2014} Although it is difficult to simultaneously reconcile all of these results, it is important to recognize that astrophysical uncertainties and the full set of nuclear operators in an effective field theory treatment of WIMP-nucleon scattering can greatly reduce the tension between null results and signals in experiments that use different target elements.\cite{Haxton2013,Gondolo2014} For this reason, it is extremely valuable to continue probing the low mass region with a variety of target elements. Ge is a particularly interesting target element because it is used by both SuperCDMS and CoGeNT, so a comparison of results from the two experiments is fully model-independent.

\section{The SuperCDMS Experiment and Detectors}
SuperCDMS at Soudan is an upgrade to the CDMS II experiment, which used similar cryogenic Ge detectors instrumented with phonon and ionization sensors. The ratio of the ionization and phonon energies, called the ``ionization yield," is a powerful discriminator of the electron recoils caused by background gammas from the nuclear recoils caused by a WIMP signal. SuperCDMS uses the same low-background experimental setup as CDMS II at the Soudan Underground Laboratory in Minnesota, but features 600~g detectors instead of the 250~g Ge detectors previously used. In total, the array has 15 detectors with a total target mass of 9~kg.

The larger SuperCDMS detectors, known as iZIPs, employ a new sensor layout which provides improved rejection of events near the surfaces of the detector. These ``surface events" have a reduced ionization yield that can mimic nuclear recoils and were the limiting background of the CDMS II experiment. Ionization and phonon sensors are photolithographically deposited in an interleaved pattern on the top and bottom surfaces of each cylindrical detector. Ionization electrodes are held at +2~V (-2~V) on the top (bottom) face of each detector, while phonon sensors are at 0~V. This configuration induces a signal in the ionization readout on both sides when an event is in the bulk of the detector, but readout on only one side for events near the surface. Dedicated calibrations with a $^{210}$Pb surface event source have measured the background leakage fraction of surface events on the top and bottom faces to be $<1.7 \times 10^{-5}$ at 90\% C.L. over 8-115~keV recoil energy with 50\% nuclear recoil acceptance.\cite{iZIP2013} An outer guard ionization electrode additionally rejects events that interact in the outer sidewalls.

\begin{figure}
\begin{minipage}{0.5\linewidth}
\centerline{\includegraphics[width=1.0\linewidth]{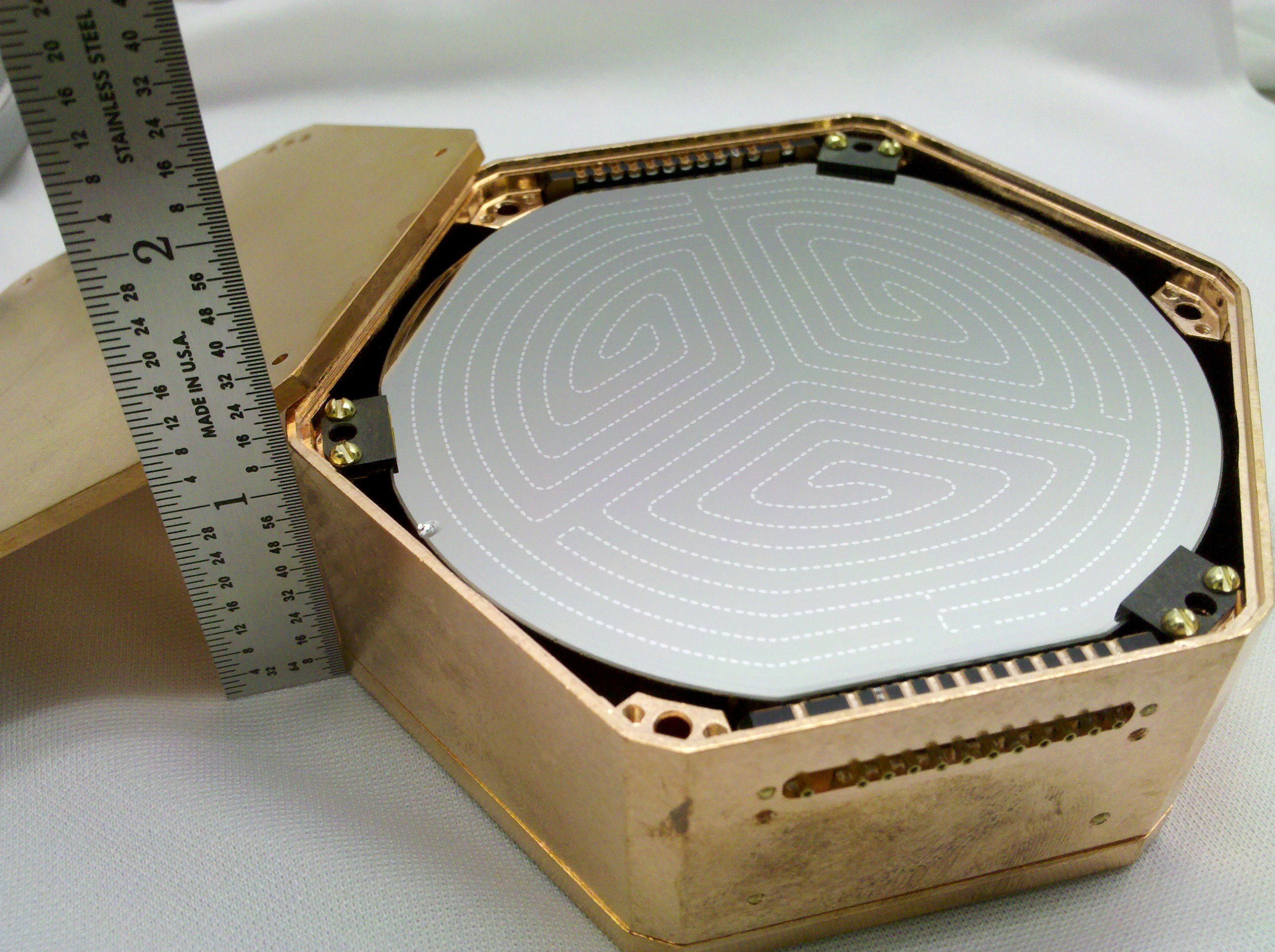}}
\end{minipage}
\hfill
\begin{minipage}{0.5\linewidth}
\centerline{\includegraphics[width=1.0\linewidth]{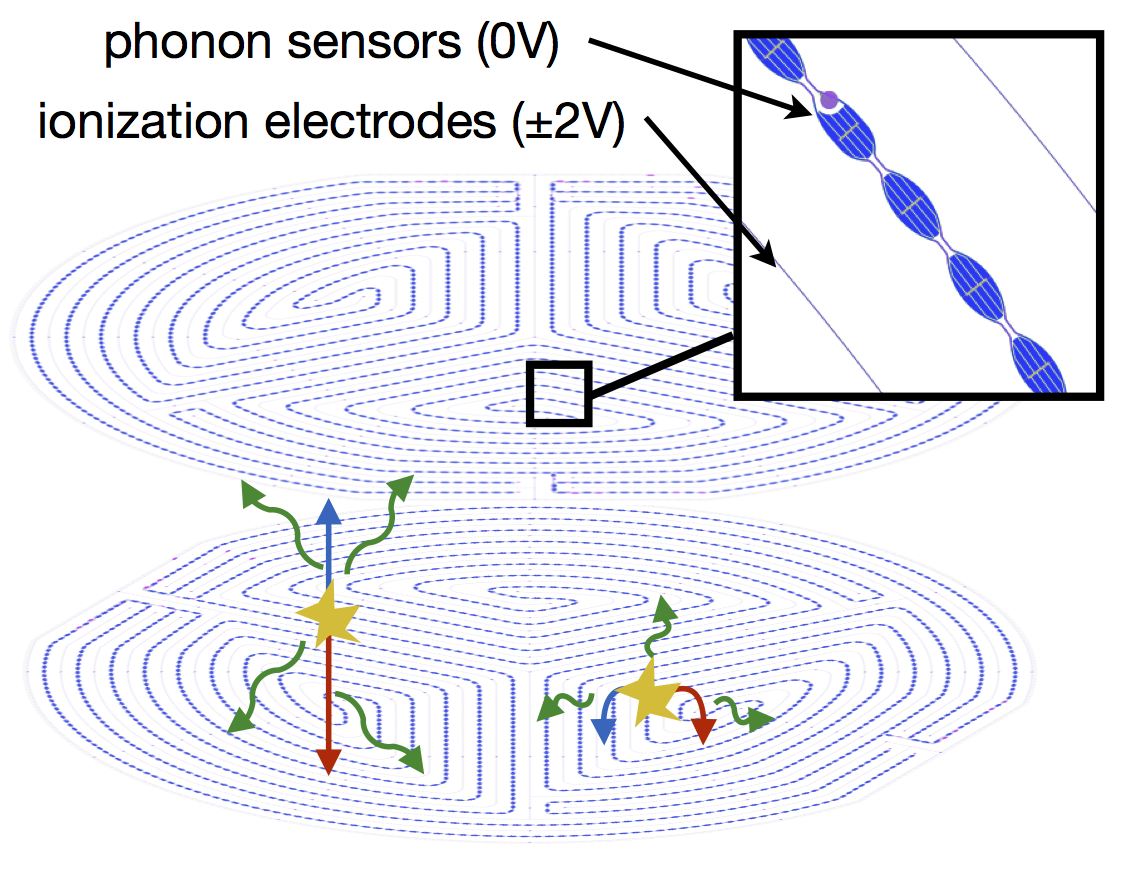}}
\end{minipage}
\caption[]{\emph{Left panel:} SuperCDMS iZIP detector mounted in housing, showing the pattern of phonon and ionization sensors photolithographically patterned on the top surface. \emph{Right panel:} Diagram showing the electron/hole (red/blue) propagation for bulk (left) and surface (right) events. Phonons (green) are produced by the initial recoil and charge carrier propagation (Neganov-Luke phonons) and provide some sensitivity to the event position. Inset shows a magnified image of the interleaved charge and phonon sensors on the top and bottom surfaces.}
\label{fig:detector}
\end{figure}

Position information is also contained in the phonon signal of the iZIP.\cite{Anderson2014} The phonon readout is fast enough that the sensors are able to measure non-equilibrium phonons before they downconvert and become uniform throughout the crystal. Positioning is possible because phonon sensors are partitioned into four channels on each face of the crystal, one of which is an outer guard ring for rejecting events on the sidewalls. Charge carriers produce Luke-Neganov phonons as they propagate through the crystal.\cite{Luke1988} Similar to the ionization, Luke-Neganov phonons produced by charges from surface events tend to produce a larger signal on one side of the detector, while events in the bulk have similar phonon signal on both sides. This position-dependence of the phonons is essential for probing low energy recoils: because of quenching, only $\sim$15-30\% of the recoil energy of a nuclear recoil is converted into ionization,\cite{Lindhard1963} while the \emph{full} recoil energy is recovered in phonons with negligible quenching. In combination with the better phonon baseline resolution in iZIP detectors (200~eV vs. 400~eV for ionization), this means that phonon-based readout has much better signal-to-noise at energies below about 5 keVnr, where the ionization readout loses sufficient resolution to be useful.

\section{Constraints on Light WIMPs}
For heavy nuclei used in dark matter searches, such as Ge or Xe, lighter WIMPs produce lower-energy recoils. To improve the sensitivity at low WIMP masses, an experiment can therefore lower its energy threshold to integrate more of the WIMP recoil spectrum, and/or it can reduce the background at the lowest energies to which it is sensitive. SuperCDMS has used both approaches to improve light WIMP exclusions in two recent analyses.

\subsection{CDMSlite}
By applying a high bias voltage ($\sim 70$~V) to a detector, the effective energy threshold of the experiment can be significantly reduced by exploiting the Luke-Neganov effect. Luke-Neganov phonons are produced in proportion to the number of charge carriers in each event
\begin{equation}
E_{Luke} = N_{eh} e V_b,
\end{equation}
where $N_{eh}$ is the number of charge carriers and $V_b$ is the bias voltage across the crystal. As the bias is increased, the phonon readout therefore acts as an amplifier of the ionization signal, allowing lower energies to be measured with a fixed hardware threshold. The trade-off of this high-voltage mode, called CDMSlite,\cite{CDMSlite2014} is that electron and nuclear recoils can no longer be discriminated because the phonon readout is used to measure ionization: the energy threshold is lowered but the background discrimination based on ionization yield is also lost.

\begin{figure}
\begin{minipage}{1.0\linewidth}
\centerline{\includegraphics[width=0.5\linewidth]{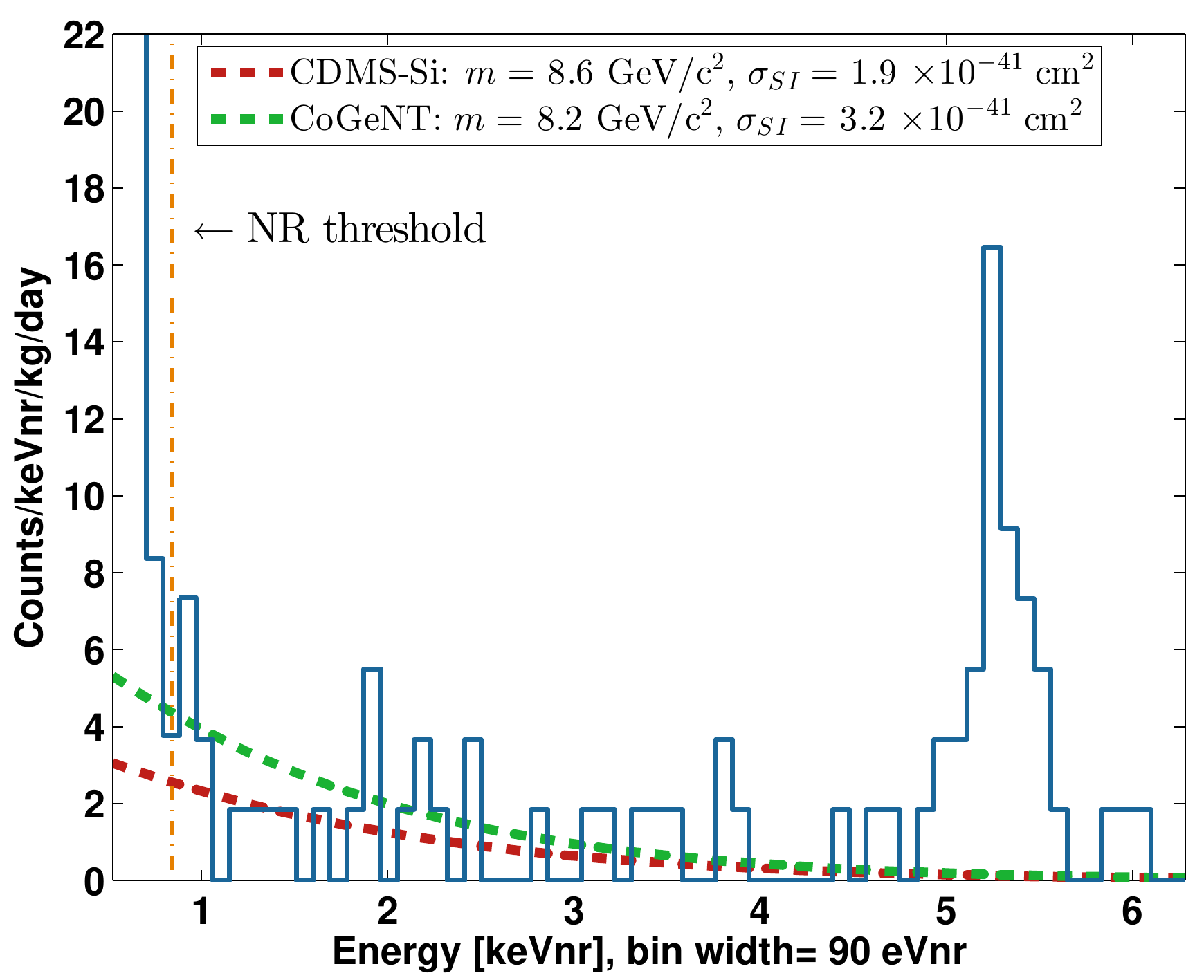}}
\end{minipage}\caption[]{Energy spectrum of events in the 6~kg-d exposure of CDMSlite (taken from ref. \cite{CDMSlite2014}).}
\label{fig:cdmslite}
\end{figure}

As a proof of concept of Luke-Neganov amplification in a dark matter experiment, an exposure of 6~kg-d was acquired on a single detector operating in the CDMSlite mode. A bias voltage of 69~V was used, corresponding to a factor 24x amplification of the ionization readout (limited to 12x due to readout electronics). With this voltage, an analysis threshold of 170~eVee or 840~keVnr (for $k=0.157$ Lindhard energy scale) is achieved (100\% efficient above threshold). The average background rate in the energy range of 0.2 and 1~keVee (0.8 to 4.2~keVnr) relevant to light WIMP interpretations of CDMS II (Si) and CoGeNT is $5.2 \pm 1$ counts/keVee/kg-d. Because of the lack of background discrimination in this energy range, a conservative 90\% C.L. upper limit set without background subtraction requires the WIMP-nucleon cross section to be $<3.5 \times 10^{-41}$~\csunit~at 8~\gev. The energy spectrum of the data, shown in Figure \ref{fig:cdmslite}, appears to be in moderate tension with the CDMS II (Si) and CoGeNT best-fit WIMP models, but the conservative upper limit cannot completely exclude the parameter space associated with those analyses because background subtraction is not used in the CDMSlite limit. Acquisition of larger exposures is ongoing in order to better characterize the background and improve the analysis techniques.

\subsection{Low-energy analysis of SuperCDMS data}
The large exposure of data taken with iZIPs operated in the normal bias configuration ($\pm 2$~V) can also stringently constrain light WIMPs because of the position-sensitivity and background rejection of the detectors.\cite{SuperCDMS2014} The position sensitivity of the phonon readout can be used to reject backgrounds from passive material surrounding the detectors, which preferentially interact in the detector surfaces. And the simultaneous measurement of ionization and phonon energy in this mode can reduce much of the background from electron recoils in the bulk of each detector.

Data from seven detectors with low energy thresholds were analyzed, totaling 577~kg-d of exposure after removing periods of poor or abnormal detector performance. Trigger and analysis energy thresholds vary between detectors and as a function of time, but extend as low as 1.5~keVnr. Events consistent with nuclear recoils were blinded and not analyzed until the final event selection was determined.

There are three primary backgrounds at low-energies in SuperCDMS: gammas from radioactivity in the passive material surrounding the detectors, L-shell electron-capture lines near 1~keVee, and surface events from $^{210}$Pb decay products that primarily originate in the copper housings surrounding the sidewalls of the detectors. The energy spectra for each background source was determined using Geant4 simulations of the contamination and decay chains, while the detector response in the ionization and phonon channels was modeled using a data-driven pulse simulation. The various backgrounds can be efficiently separated in high-energy sidebands ($>10$~keVee) where the signal-to-noise is excellent. Events from calibration and low-background data in these sidebands were treated as templates of each background type and summed with randomly sampled noise to mimic the response of the detector. 

The event selection for this search consists of three tiers of requirements. The first tier removes poorly reconstructed and noise-induced events. The second tier requires that events be consistent with WIMPs in ionization and phonon topology, as well as removing events that scatter in multiple detectors, which are unlikely to be WIMPs. The third tier of event selection uses a boosted decision tree (BDT) to optimize the 90\% C.L. upper limit on the WIMP-nucleon cross section, given the model of known backgrounds and the detector response.

The BDT was trained on four discriminator variables chosen for their ability to tag each type of background: ionization energy, total phonon energy, radial phonon partitioning, and $z$ phonon partitioning. The latter two variables correspond respectively to the fraction of energy measured by the outer phonon channels and the fraction of energy measured by the phonon channels on the top face versus the bottom face. The ratio between the ionization and phonon energies can identify electron recoils, while the phonon partition information can identify the surface events from $^{210}$Pb decay products. Each BDT reduces the 4-dimensional parameter space to a single discriminant variable. The acceptance region of the BDT output parameter was set for each WIMP mass and detector by simultaneously optimizing the expected 90\% C.L. upper limit on the WIMP-nucleon cross section. Separate BDTs were trained for each detector to discriminate the backgrounds from the signal produced by a 5, 7, 10, and 15~\gev~WIMPs. The final event selection was taken to be the logical OR of the BDTs for each WIMP mass---an \emph{ad hoc} technique that provides good sensitivity to a broad range of WIMP masses when used with the optimal interval method\cite{Yellin2002} for setting the upper limit.

\begin{figure}
\begin{minipage}{0.5\linewidth}
\centerline{\includegraphics[width=1.0\linewidth]{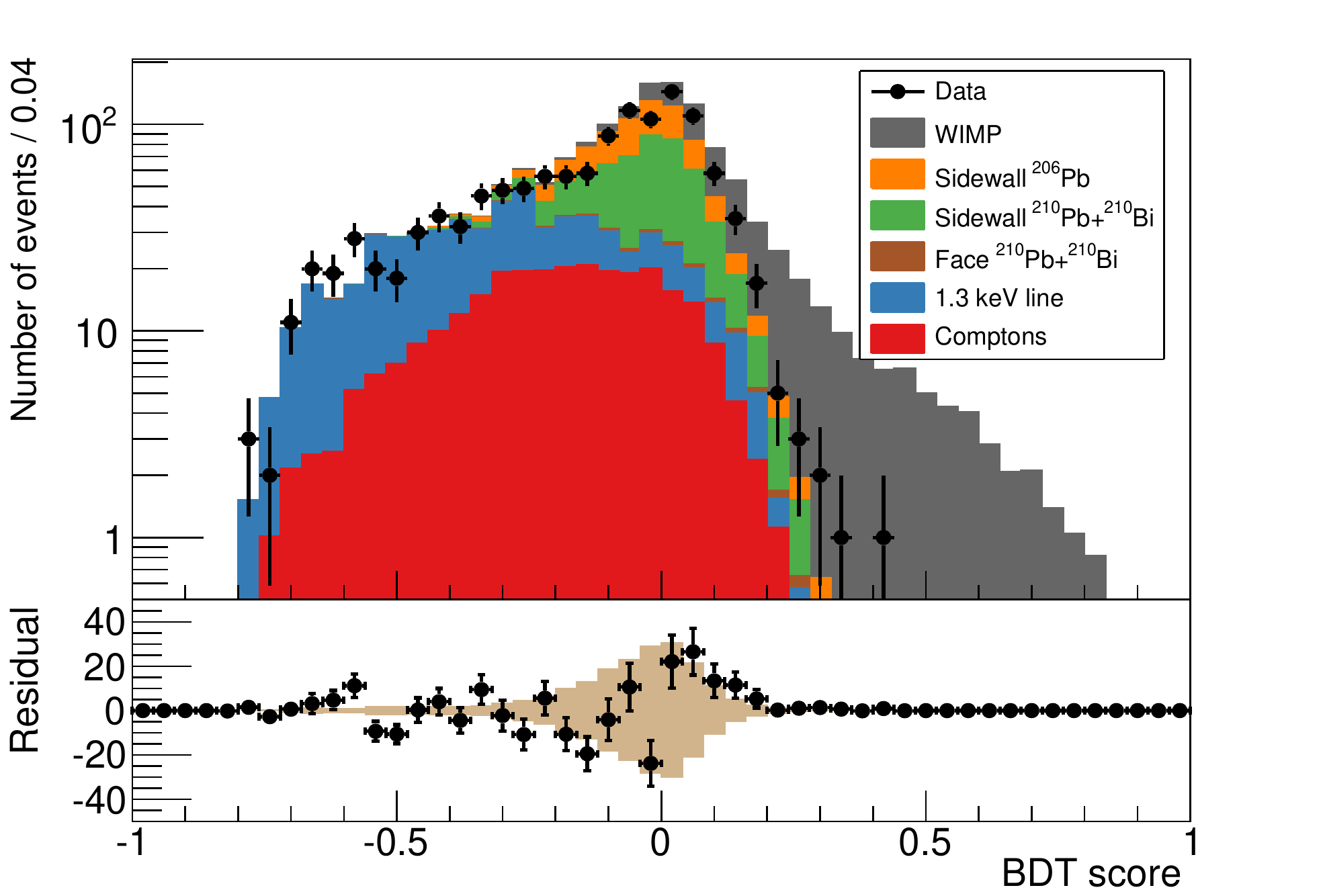}}
\end{minipage}
\hfill
\begin{minipage}{0.5\linewidth}
\centerline{\includegraphics[width=1.0\linewidth]{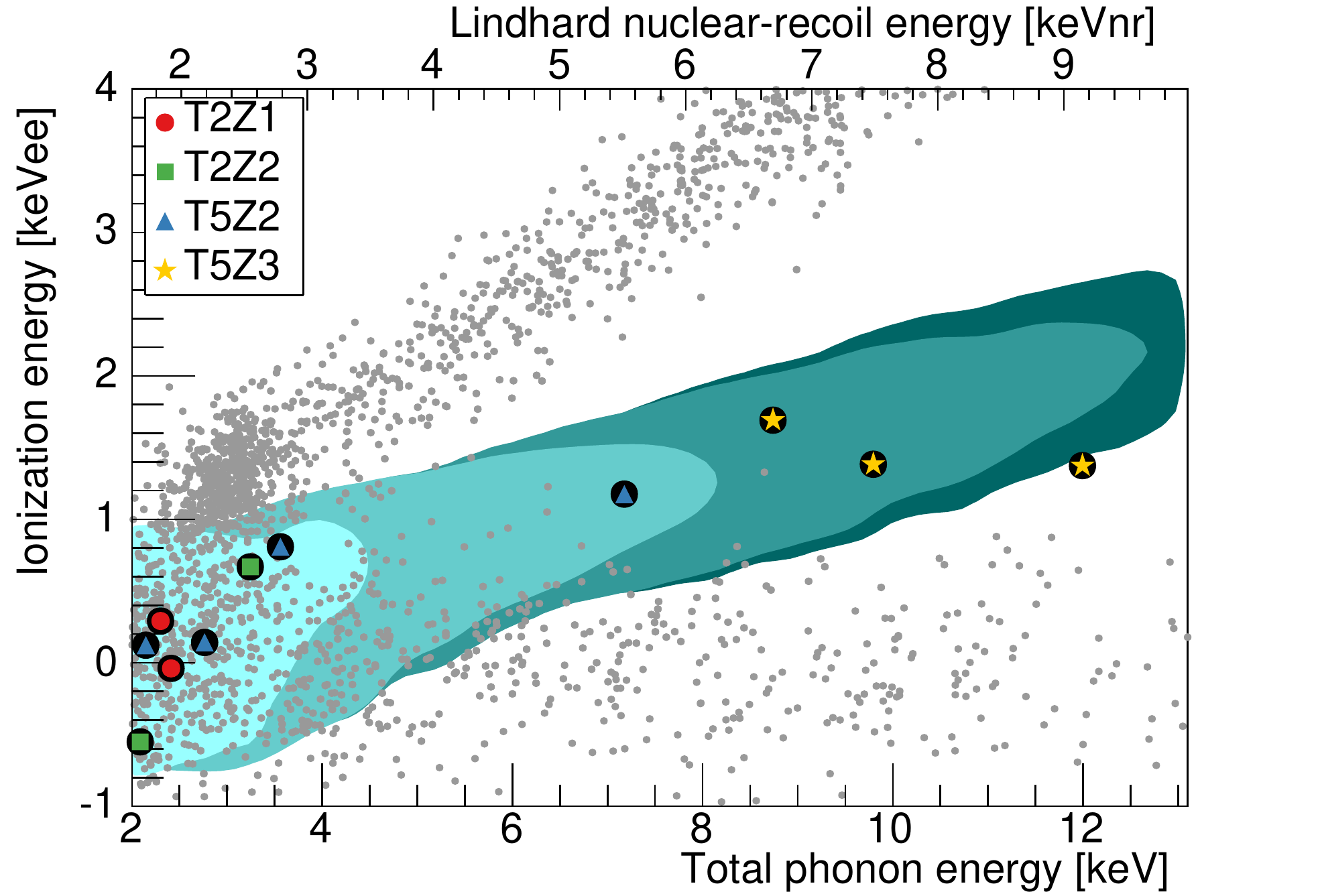}}
\end{minipage}
\caption[]{\emph{Left panel:} BDT output for data (\emph{black}), background model components (\emph{stacked colors}), and a 10~\gev~WIMP ($\sigma = 6 \times 10^{-42}$~\csunit) summed across the seven detectors used in the low-energy analysis of SuperCDMS data. Bottom panel shows the residual between the data and background model, with the background model systematic error shown in tan bars. \emph{Right panel:} Grey points are all single-scatter events within the ionization fiducial volume that pass data quality selections. Large encircled shapes are the 11 candidate events passing the BDT selection, labeled by detector. Shaded regions correspond to 95\% contours for 5, 7, 10, and 15~\gev~WIMPs after applying all selection criteria (taken from ref. \cite{SuperCDMS2014}).}
\label{fig:scdmsdata}
\end{figure}

Before unblinding, $1190 \pm 85$ events were expected to pass all selection criteria before the BDT, while 6.2 + 1.1 - 0.8 were expected also to pass the BDT event selection. After unblinding, 1218 events pass all criteria before the BDT, and 11 events pass the BDT, as shown in Figure \ref{fig:scdmsdata}. Including systematic errors on the background model, the $p$-value for observing 11 or more events is 0.07, so no evidence of a WIMP signal can be claimed. Within the region of events consistent with WIMPs, this demonstrates a background rejection of 99.1\% at a spectrum-averaged exposure of 10\% for a 10~GeV WIMP (signal acceptance of 22\% above threshold).

Observed background rates are generally consistent with expectations on all detectors, except for one detector called T5Z3, which has three events while only 0.13 + 0.06 - 0.04 are expected. These three events are at energies significantly above those expected from known background sources. This detector has a shorted ionization guard electrode which significantly modifies its electric field, and although the background model correctly accounted for the lack of data from this channel, we discovered \emph{after unblinding} that the modified electric field biases the selection of template events causing some mismodeling of the background. Detailed simulations and calibrations of different electric field configurations are in progress to better understand these effects.

\begin{figure}
\begin{minipage}{1.0\linewidth}
\centerline{\includegraphics[width=0.6\linewidth]{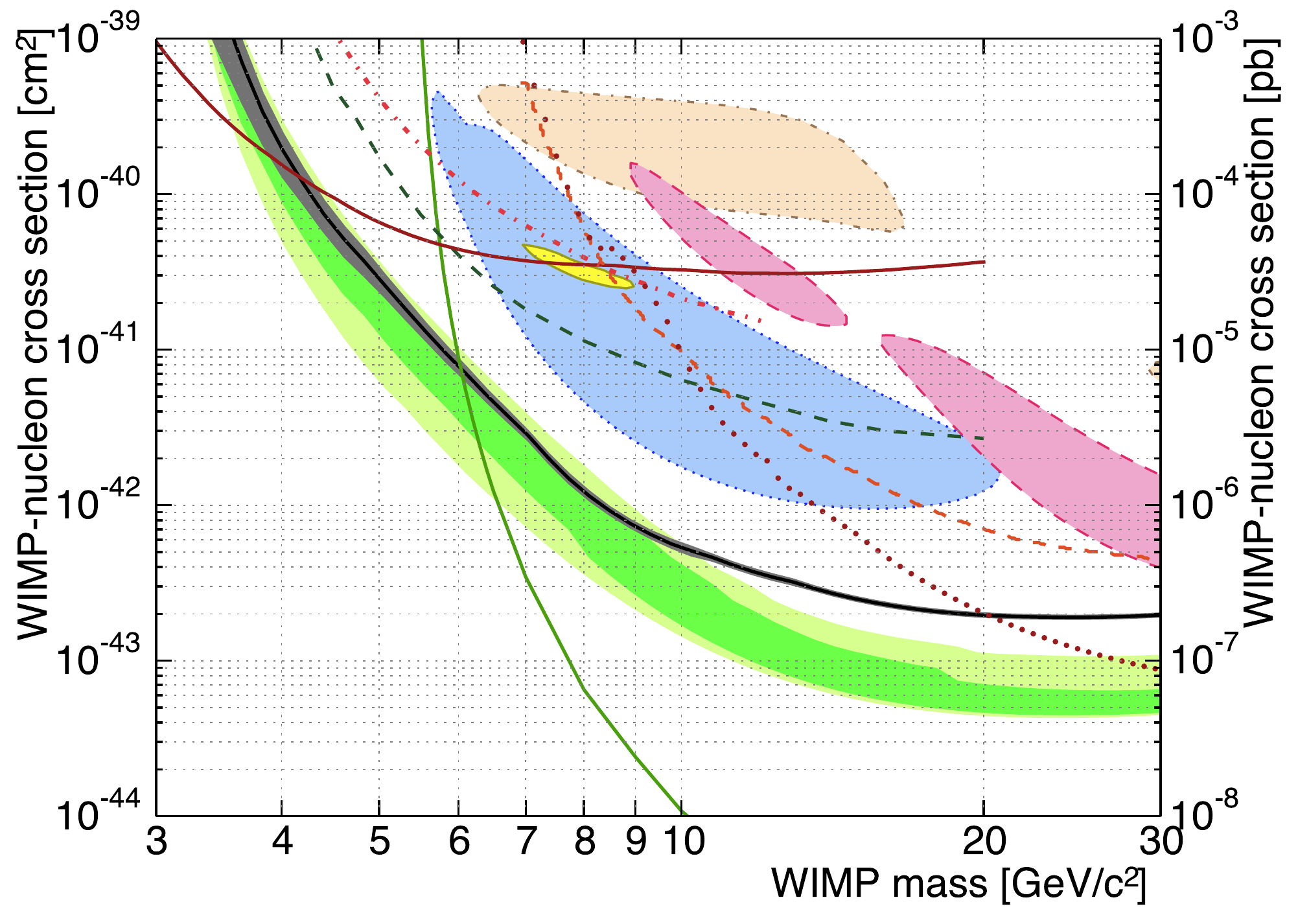}}
\end{minipage}\caption[]{The 90\% confidence upper limit (solid black) based on all observed events is shown with 95\% C.L. systematic uncertainty band (gray).  The pre-unblinding expected sensitivity in the absence of a signal is shown as 68\% (dark green) and 95\% (light green) C.L. bands.  The disagreement between the limit and sensitivity at high WIMP mass is due to the events in T5Z3.  Closed contours shown are CDMS II (Si)\cite{CDMSSi2013} (\emph{dotted blue}, 90\% C.L.), CoGeNT\cite{CoGeNT2013} (\emph{yellow}, 90\% C.L.), CRESST\cite{CRESST2012} (\emph{dashed pink}, 95\% C.L.), and DAMA/LIBRA\cite{DAMA2013} (\emph{dash-dotted tan}, 90\% C.L.).  90\% C.L. exclusion limits shown are CDMS II (Ge)\cite{CDMSIIGe2010} (\emph{dotted dark red}), CDMS II (Ge) low-threshold\cite{CDMSGe2011} (\emph{dashed-dotted red}), CDMSlite\cite{CDMSlite2014} (\emph{solid dark red}), LUX\cite{LUX2014} (\emph{solid green}), XENON10 S2-only\cite{XENON102011,XENON102011err} (\emph{dashed dark green}), and EDELWEISS low-threshold\cite{EDELWEISS2012} (\emph{dashed orange}) (taken from ref. \cite{SuperCDMS2014}).}
\label{fig:limits}
\end{figure}

Conservatively treating the 11 events as potential WIMPs, a 90\% C.L. upper limit was computed using the optimal interval method~\cite{Yellin2002} and standard halo parameters,\cite{CDMSIIiDM2011} shown in Figure \ref{fig:limits}. The limit excludes WIMP interpretations of CDMS II (Si), DAMA/LIBRA, and CRESST under standard halo assumptions. Because both SuperCDMS and CoGeNT have a Ge target, this result furthermore excludes the CoGeNT result in a model-independent fashion. A WIMP at the CoGeNT best-fit mass and cross section would produce approximately 210 events passing the BDT event selection in SuperCDMS. In addition, the SuperCDMS low-energy search significantly improves the sensitivity of non-Xe based experiments in the 4-20~\gev~mass range.

\section{Conclusions and Outlook}
The combination of the SuperCDMS dedicated low-mass WIMP searches and LUX have now severely constrained WIMP interpretations of dark matter anomalies at low masses. Significant improvement in the sensitivity of the CDMSlite approach may be possible by using background subtraction and exploiting the position-sensitivity of the iZIP to Luke-Neganov phonons. Improved pulse reconstruction algorithms\cite{Hertel2012,Schlupf2014} and refinements to the background modeling also should provide significant increases in sensitivity for low-energy WIMP searches in the normal bias configuration. The SuperCDMS approaches have demonstrated sensitivity to very low recoil energies as well as background rejection at the $>99$\% level in the 1.5 to 10 keVnr energy range. These two analyses also introduce the method of Luke-Neganov amplification and BDTs into direct dark matter searches for the first time. When combined with larger experimental masses and well-understood improvements in the materials screening planned for a proposed 100~kg experiment at SNOLAB, these methods will continue to advance sensitivity at WIMP masses near a few \gev. Given the large available parameter space, it is essential to continue to explore the low-mass region with multiple target materials in the search for WIMPs.

\section*{Acknowledgments}

I thank Julien Billard for valuable discussions and acknowledge support from the Department of Energy Office of Science Graduate Fellowship Program (DOE SCGF), made possible in part by the American Recovery and Reinvestment Act of 2009, administered by ORISE-ORAU under contract no. DEAC05-06OR23100. The SuperCDMS collaboration acknowledges technical assistance from the Soudan Underground Laboratory and Minnesota Department of Natural Resources. SuperCDMS is supported by the US DOE, NSF, NSERC Canada, and MULTIDARK.

\end{document}